\documentclass[11pt]{article}

\makeatletter

\def\paragraph{\@startsection{paragraph}{4}{\z@}{+2.00ex plus
 +1ex minus +.2ex}{1.5ex plus .2ex}{\it\normalsize}}

\def\section{\@startsection {section}{1}{\z@}{+3.0ex plus +1ex minus
  +.2ex}{2.3ex plus .2ex}{\normalsize\bf\boldmath}}
\def\subsection{\@startsection{subsection}{2}{\z@}{+2.5ex plus +1ex
minus +.2ex}{1.5ex plus .2ex}{\normalsize\bf\boldmath}}
\def\subsubsection{\@startsection{subsubsection}{3}{\z@}{+3.25ex plus
 +1ex minus +.2ex}{1.5ex plus .2ex}{\normalsize\bf\boldmath}}

\expandafter\ifx\csname mathrm\endcsname\relax\def\mathrm#1{{\rm #1}}\fi

\newcounter{saveeqn}

\@addtoreset{equation}{section}

\usepackage{graphicx,xspace}

\newcommand{\beq}{\begin{equation}}
\newcommand{\eeq}{\end{equation}}
\newcommand{\bea}{\begin{eqnarray}}
\newcommand{\eea}{\end{eqnarray}}
\newcommand{\bseq}{\begin{subequations}}
\newcommand{\eseq}{\end{subequations}}
\newcommand{\bsea}{\begin{subeqnarray}}
\newcommand{\esea}{\end{subeqnarray}}
\newcommand{\bit}{\begin{itemize}}
\newcommand{\eit}{\end{itemize}}
\newcommand{\ben}{\begin{enumerate}}
\newcommand{\een}{\end{enumerate}}
\newcommand{\bfig}{\begin{figure}}
\newcommand{\efig}{\end{figure}}
\newcommand{\btab}{\begin{table}}
\newcommand{\etab}{\end{table}}
\newcommand{\im}[1]{\ensuremath{#1}\xspace}       
\newcommand{\imx}[1]{\ensuremath{#1}\xspace}       
\newcommand{\etal}{{\em et al.}}

\newcommand{\eV}{\imx{\mathrm{e\kern -0.1em V}}}
\newcommand{\MeV}{\imx{\mathrm{Me\kern -0.1em V}}}
\newcommand{\GeV}{\imx{\mathrm{Ge\kern -0.1em V}}}
\newcommand{\TeV}{\imx{\mathrm{Te\kern -0.1em V}}}

\newcommand{\IM}{\im{\mathrm{I} \kern -0.15 em \mathrm{m}}}         
\newcommand{\RE}{\im{\mathrm{R} \kern -0.15 em \mathrm{e}}}         

\newcommand{\Pchi}{\im{{\raise5pt\hbox{$\chi$}}}}
\newcommand{\Pe}{\im{\mathrm{e}}}            
\newcommand{\Pm}{\im{\mu}}
\newcommand{\Pt}{\im{\tau}}
\newcommand{\Pn}{\im{\nu}}
\newcommand{\Pl}{\im{\ell}}
\newcommand{\Pp}{\im{\mathrm{p}}}

\newcommand{\Pc}{\im{\mathrm{c}}}
\newcommand{\Pb}{\im{\mathrm{b}}}
\newcommand{\PT}{\im{\mathrm{t}}}
\newcommand{\Pq}{\im{q}}
\newcommand{\Pf}{\im{f}}
\newcommand{\Phad}{\im{\mathrm{had}}}

\newcommand{\PW}{\im{\mathrm{W}}}             

\newcommand{\PZ}{\im{\mathrm{Z}}}
\newcommand{\PH}{\im{\mathrm{H}}}

\newcommand{\MW}{\im{M_{\PW}}}
\newcommand{\MZ}{\im{M_{\PZ}}}
\newcommand{\MH}{\im{M_{\PH}}}

\newcommand{\MT}{\im{M_{\PT}}}

\newcommand{\G}{\im{\Gamma}}                  
\newcommand{\GW}{\im{\G_{\PW}}}               

\newcommand{\A}{\im{\mathrm{A}}}

\newcommand{\Ae}{\im{\A_{\Pe}}}

\newcommand{\Al}{\im{\A_{\Pl}}}
\newcommand{\Ab}{\im{\A_{\Pb}}}

\newcommand{\Aq}{\im{\A_{\Pq}}}

\newcommand{\Afb}{\im{\A_{\mathrm{fb}}}}

\newcommand{\Afbzb}{\im{\Afb^{0,\Pb}}}
\newcommand{\Afbzc}{\im{\Afb^{0,\Pc}}}

\newcommand{\swsqeffl}{\sin^2\theta_{\mathrm{eff}}^{\mathrm{lept}}}

\newcommand{\Pee}{\im{\Pe^+\Pe^-}}
\newcommand{\Pmm}{\im{\Pm^+\Pm^-}}
\newcommand{\Ptt}{\im{\Pt^+\Pt^-}}
\newcommand{\Pll}{\im{\Pl^+\Pl^-}}
\newcommand{\Ppp}{\im{\Pp\overline{\Pp}}}

\newcommand{\Pnn}{\im{\Pn\overline{\Pn}}}
\newcommand{\Pqq}{\im{\Pq\overline{\Pq}}}
\newcommand{\PTT}{\im{\PT\overline{\PT}}}
\newcommand{\Pbb}{\im{\Pb\overline{\Pb}}}

\newcommand{\Pff}{\im{\Pf\overline{\Pf}}}
\newcommand{\PWW}{\im{\PW^+\PW^-}}
\newcommand{\PZZ}{\im{\PZ\PZ}}

\newcommand{\Peeee}{\im{\Pee \kern -0.35em \rightarrow\Pee}}
\newcommand{\Peemm}{\im{\Pee \kern -0.35em \rightarrow\Pmm}}
\newcommand{\Peett}{\im{\Pee \kern -0.35em \rightarrow\Ptt}}
\newcommand{\Peell}{\im{\Pee \kern -0.35em \rightarrow\Pll}}
\newcommand{\Peenn}{\im{\Pee \kern -0.35em \rightarrow\Pnn}}
\newcommand{\Peeqq}{\im{\Pee \kern -0.35em \rightarrow\Pqq}}
\newcommand{\Peehad}{\im{\Pee \kern -0.35em \rightarrow\Phad}}
\newcommand{\Peeff}{\im{\Pee \kern -0.35em \rightarrow\Pff}}
\newcommand{\PeeTT}{\im{\Pee \kern -0.35em \rightarrow\PTT}}
\newcommand{\PeeWW}{\im{\Pee \kern -0.35em \rightarrow\PWW}}
\newcommand{\PeeZZ}{\im{\Pee \kern -0.35em \rightarrow\PZZ}}
\newcommand{\aqcd}{\im{\alpha_S}}

\newcommand{\GF}{\im{G_{\mathrm{F}}}}

\newcommand{\dalhad}{\im{\Delta\alpha^{(5)}_{had}}}

\begin{document}

\noindent                
{\Large
 $\phantom{0}$        \hfill UCD-PHYC/051101\\[1mm]
 $\phantom{0}$        \hfill  hep-ex/0511018\\[2mm]
 $\phantom{0}$\bf     \hfill 6 November 2005\\[1mm]
}
\begin{center}

\includegraphics[width=4cm]{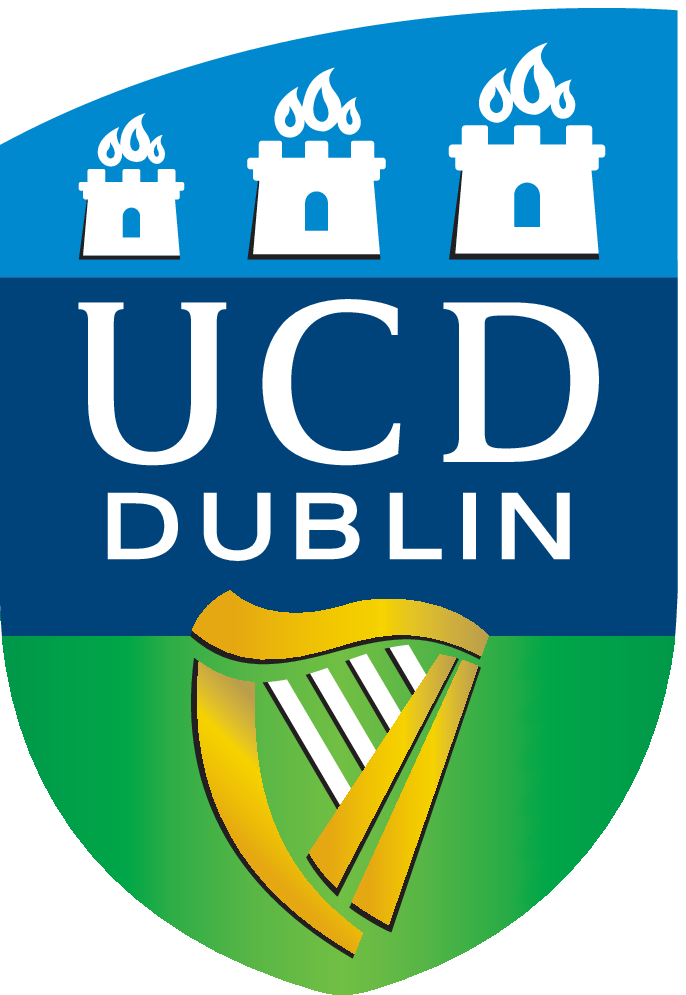}
\vskip 1cm
{\Huge\bf Precision Tests of the \\ Standard Model \\}
\vskip 1cm
{\Large {\bf Martin W. Gr\"unewald}\\[20pt]
        UCD School of Physics\\
        University College Dublin\\
        Belfield, Dublin 4\\
        Ireland\\}
\vfill
{\bf Abstract} \\[10pt]
\end{center}
{
  
  Recent published and preliminary precision electroweak measurements
are reviewed, including new results on the mass of the top quark and
mass and width of the W boson.  The experimental results are compared
with the predictions of the Standard Model and are used to constrain
its free parameters, notably the mass of the Higgs boson.  The
agreement between measurements and expectations from theory is
discussed.

}
\vskip 1cm
\begin{center}
\em{ Invited talk presented at the EPS HEPP conference,\\
  Lisboa, Portugal, July 21st to 27th, 2005}
\end{center}

\clearpage

\setcounter{page}{2}

\section{Introduction}

\vskip -2mm On the level of realistic observables such as measured
cross sections, ratios and asymmetries, the electroweak precision data
consist of over thousand measurements with partially correlated
statistical and systematic uncertainties. This large set of results is
reduced to a more manageable set of 17 precision results, so-called
pseudo observables, in a largely model-independent procedure, by the
LEP and Tevatron Electroweak Working Groups.  The pseudo observables
updated for this conference are briefly reviewed and Standard Model
analyses~\cite{LEPEWWG} are performed, where the hadronic vacuum
polarisation at the Z pole and ``constants'' such as the Fermi
constant $\GF$ are used as well.

\section{Measurements}

\vskip -2mm
About 3/4 of all pseudo observables arise from measurements performed
in electron-positron collisions at the Z resonance, by the SLD
experiment and the LEP experiments ALEPH, DELPHI, L3 and OPAL.  The
Z-pole observables are: 5 observables describing the Z lineshape and
leptonic forward-backward asymmetries, 2 observables describing
polarised leptonic asymmetries measured by SLD with polarised beams
and at LEP exploiting tau polarisation, 6 observables describing b-
and c-quark production at the Z pole, and finally the inclusive
hadronic charge asymmetry.  The Z-pole results and their combinations
are final and by now published~\cite{Z-POLE}.  The remaining pseudo
observables are: the mass and total width of the W boson measured by
CDF and {D\O} at the Tevatron and by the four LEP-II experiments, and
the top quark mass measured at the Tevatron.
\vfill
{\bf Mass of the Top Quark}

Ten years ago the Tevatron experiments CDF and {D\O} discovered the
top quark in proton-antiproton collisions at $1.8~\TeV$ centre-of-mass
energy, by observing the reaction $\Ppp\to\PTT~X,~\PTT\to\Pbb\PWW$.
The published results based on data collected during Run-I (1992-1996)
and the preliminary results based on Run-II data (since 2001)
presented at this conference~\cite{Mtop-EPS} are combined by the
Tevatron Electroweak Working~\cite{TEVEWWG05-EPS}: $\MT =
172.7\pm1.7~(stat.)\pm2.4~(syst.)~\GeV$.
\vfill
{\bf Mass and Width of the W boson}

Final results on $\MW$ and $\GW$ from CDF and {D\O} are available for
the complete Run-I data set and are combined taking correlations
properly into account~\cite{TEV-MW-GW}.  No results are available for
Run-II data yet.  The results from ALEPH, DELPHI and L3 are
preliminary, while OPAL has recently published final results for their
complete LEP-II data set~\cite{MW-EPS}.  The combined results of the
Tevatron ($\MW=80.452\pm0.059~\GeV$) and LEP-II
($\MW=80.392\pm0.039~\GeV$) are in very good agreement.
\vfill
{\bf Z Decays to b and c Quarks}

\begin{figure}[p]
\begin{center}
$ $\vskip -1.2cm
\includegraphics[width=0.4\linewidth]{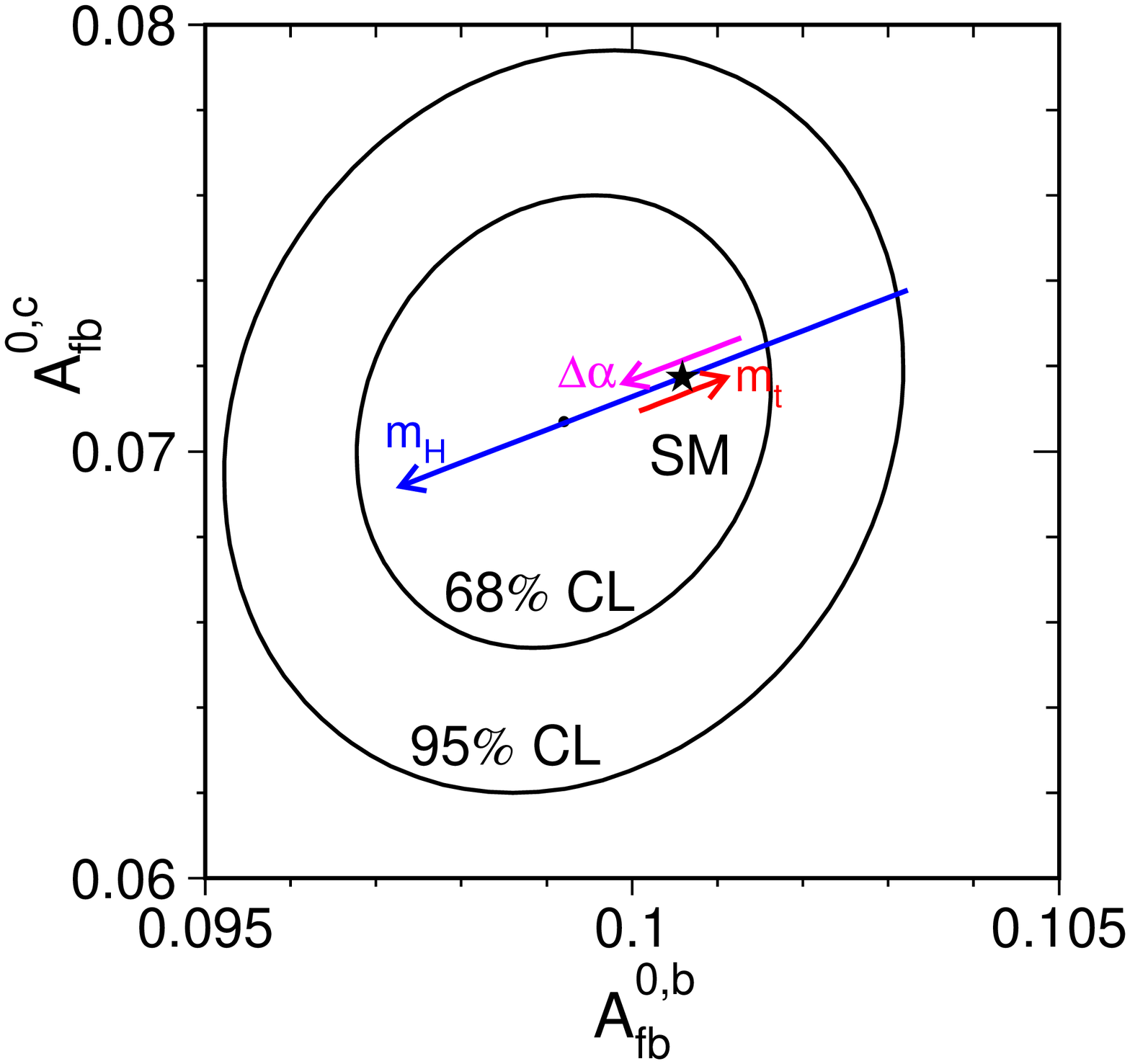}
\hfill
\includegraphics[width=0.4\linewidth]{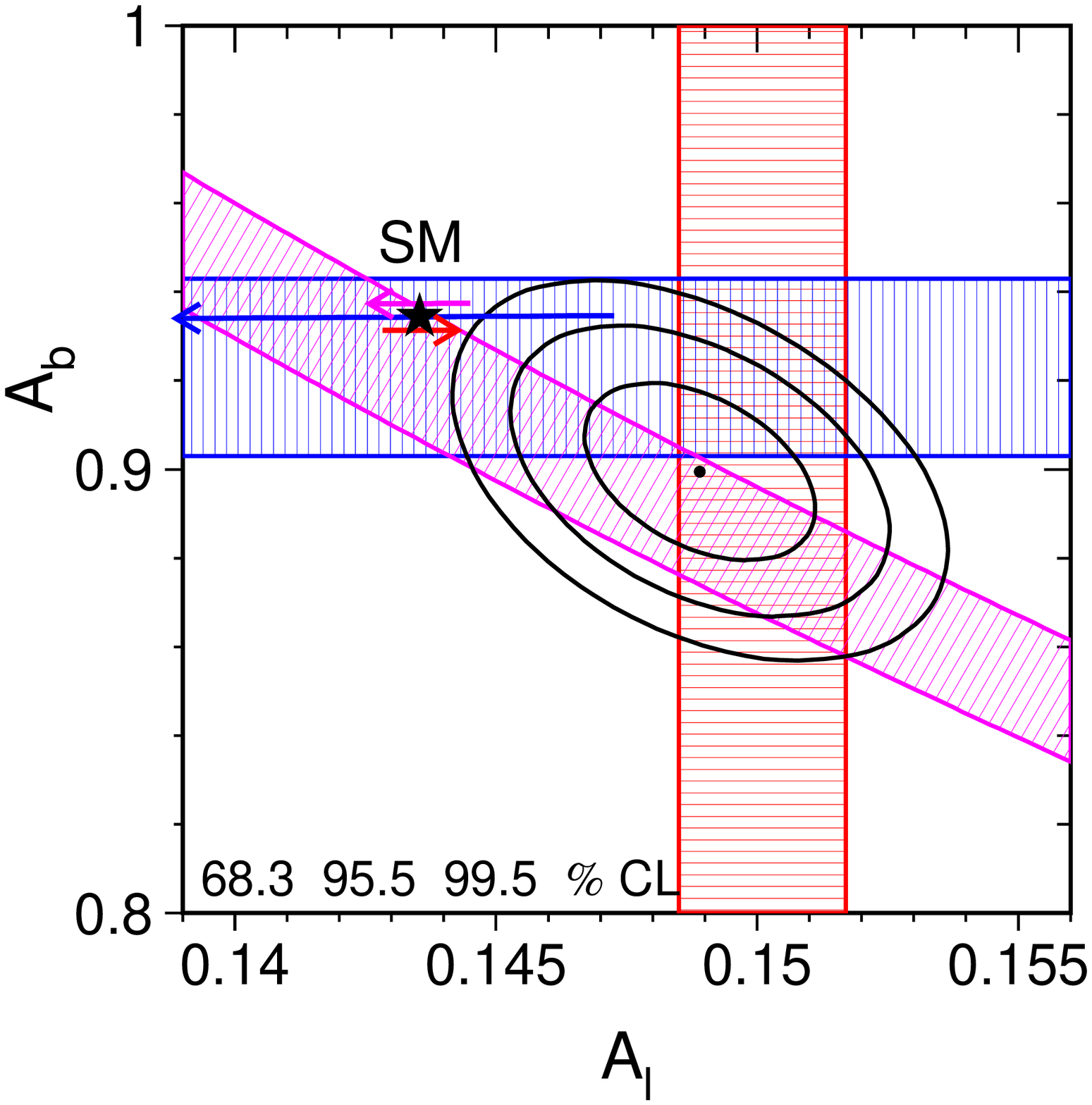}
\vskip -0.9cm
\caption{Left: Contour curves in the $(\Afbzb,\Afbzc)$ plane.  Right:
  Bands of $\pm1\sigma$ showing the combined results of $\Al$, $\Ab$,
  and $\Afbzb=3/4\Ae\Ab$.  The SM expectations are shown as the arrows
  for $\MT=172.7\pm2.9~\GeV$ and $\MH=300^{+700}_{-186}~\GeV$ and
  $\dalhad=0.02758\pm0.00035$.}
\label{fig:coup:aq}
\end{center}
\end{figure}

The heavy-flavour results at the Z-pole were the last precision
electroweak Z-pole results to become final.  Details on the various
heavy-flavour measurements at the Z pole are given in~\cite{Z-POLE}.
The combination has a rather low $\chi^2$ of 53 for $(105-14)$ degrees
of freedom: all forward-backward asymmetries are very consistent, and
their combination is still statistics limited.  The combined values
for $\Afbzb$ and $\Afbzc$ are compared to the SM expectation in
Figure~\ref{fig:coup:aq} (left), showing that $\Afbzb$ agrees well
with the SM expectation for an intermediate Higgs-boson mass of a few
hundred $\GeV$.  The mutual consistency of the measurements of $\Ab$,
$\Afbzb=(3/4)\Ae\Ab$ and $\Al$ assuming lepton universality is shown
in Figure~\ref{fig:coup:aq} (right).  Compared to the experimental
uncertainties, the SM predictions are nearly constant in $\Aq$, in
contrast to the situation for $\Al$.  This is a consequence of the SM
values of electric charge and iso-spin of quarks.

\clearpage

{\bf Effective Electroweak Mixing Angle}

Assuming the SM structure of the effective coupling constants, the
measurements of the various asymmetries are compared in terms of
$\swsqeffl$ in Figure~\ref{fig:sef2-mt-mw} (left).  The average of all
six $\swsqeffl$ determinations is $\swsqeffl = 0.23153\pm0.00016$,
with a $\chi^2/dof$ of 11.8/5, corresponding to a probability of
3.7\%. The enlarged $\chi^2/dof$ is solely driven by the two most
precise determinations of $\swsqeffl$, namely those derived from the
measurements of $\Al$ by SLD, dominated by the left-right asymmetry
result, and of $\Afbzb$ at LEP.  These two measurements differ by 3.2
standard deviations.  This is a consequence of the same effect as
shown in Figure~\ref{fig:coup:aq} (right).

\section{Global Standard Model Analysis}
\label{sec:MSM}

\vskip -2mm 
Within the framework of the SM, each pseudo observable is calculated
as a function of five main relevant parameters, which are the running
electromagnetic and strong coupling constant evaluated at the Z pole,
$\alpha_{em}$ and $\aqcd$, and the masses of Z boson, top quark and
Higgs boson, $\MZ$, $\MT$, $\MH$.  Using the Fermi constant $\GF$
allows to calculate the mass of the W boson. The running
electromagnetic coupling is represented by the hadronic vacuum
polarisation $\dalhad$, as it is this contribution which has the
largest uncertainty, $\dalhad=0.02758\pm0.00035$~\cite{BP05}.  The
precision of the Z-pole measurements requires matching precision of
the theoretical calculations. The dependence on $\MT$ and $\MH$ enters
through radiative corrections.  The predictions are calculated with
the computer programs~\cite{TZ} TOPAZ0 and ZFITTER, which incorporate
state-of-the-art calculations.

\begin{figure}[p]
\begin{center}
$ $\vskip -1cm
\includegraphics[width=0.34\linewidth]{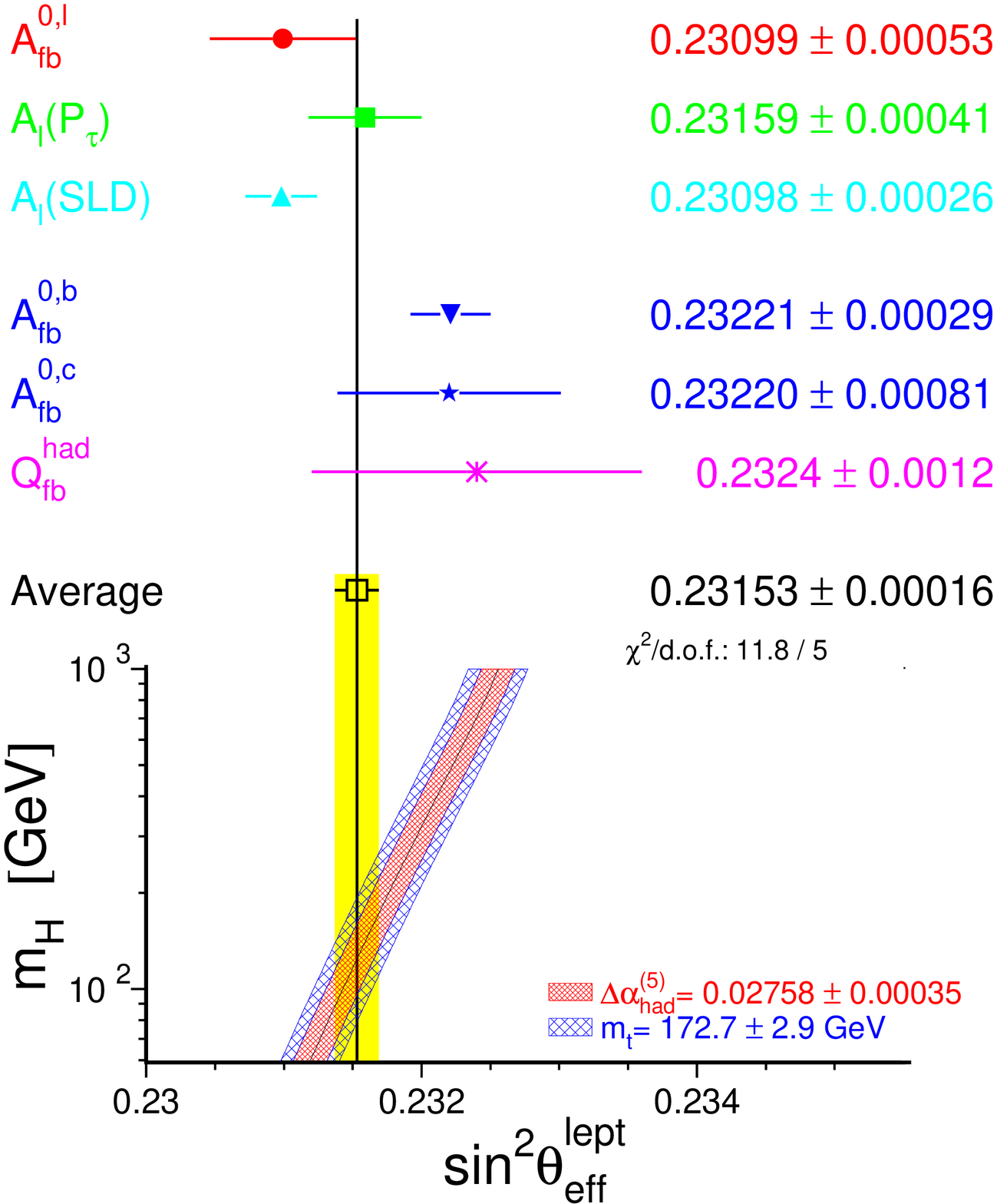}
\hfill
\includegraphics[width=0.4\linewidth]{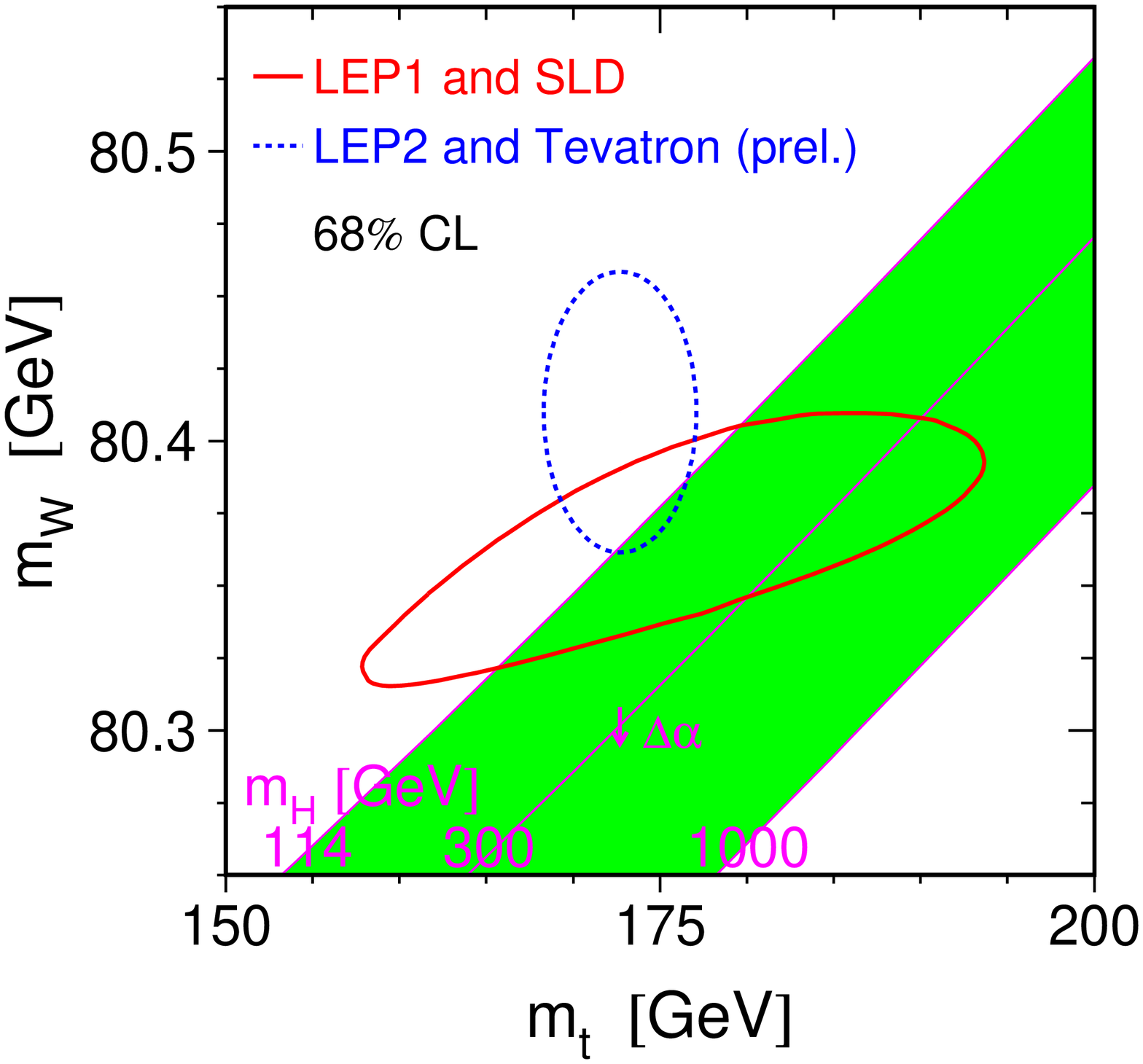}
\vskip -0.5cm
\caption{Left: The effective electroweak mixing angle from asymmetry
  measurements. Right: Contour curves of 68\% C.L. in the $(\MT,\MW)$
  plane for the direct measurements and the indirect determinations.
  The band shows the correlation between $\MW$ and $\MT$ expected in
  the SM. }
\label{fig:sef2-mt-mw}
\end{center}
\end{figure}

Using the Z-pole measurements of SLD and LEP-I in order to evaluate
electroweak radiative corrections, the masses of the two heavy
particles, the top quark and the W boson, can be predicted. The
resulting 68\% C.L. contour curve in the $(\MT,\MW)$ plane is shown in
Figure~\ref{fig:sef2-mt-mw} (right).  Also shown is the contour curve
corresponding to the direct measurements of both quantities at the
Tevatron and at LEP-II. The two contours overlap, successfully testing
the SM at the level of electroweak radiative corrections. The diagonal
band in Figure~\ref{fig:sef2-mt-mw} (right) shows the constraint
between the two masses within the SM, which depends on the mass of the
Higgs boson, and to a small extent also on the hadronic vacuum
polarisation (small arrow labeled $\Delta\alpha$).  Both the direct
and the indirect contour curves prefer a low value for the mass of the
SM Higgs boson.

\begin{figure}[p]
\begin{center}
$ $\vskip -1cm
\includegraphics[width=0.3\linewidth]{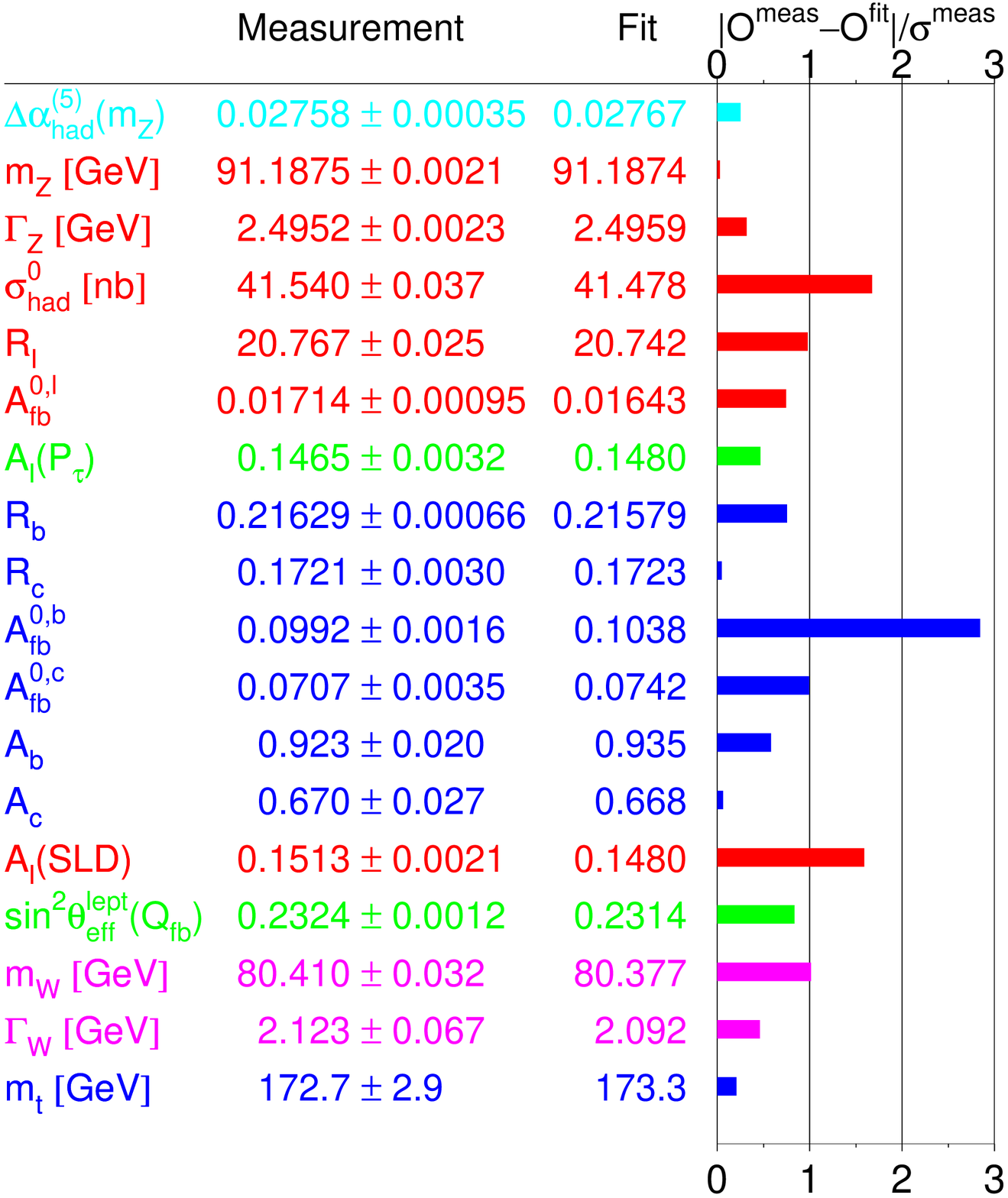}
\hfill
\includegraphics[width=0.4\linewidth]{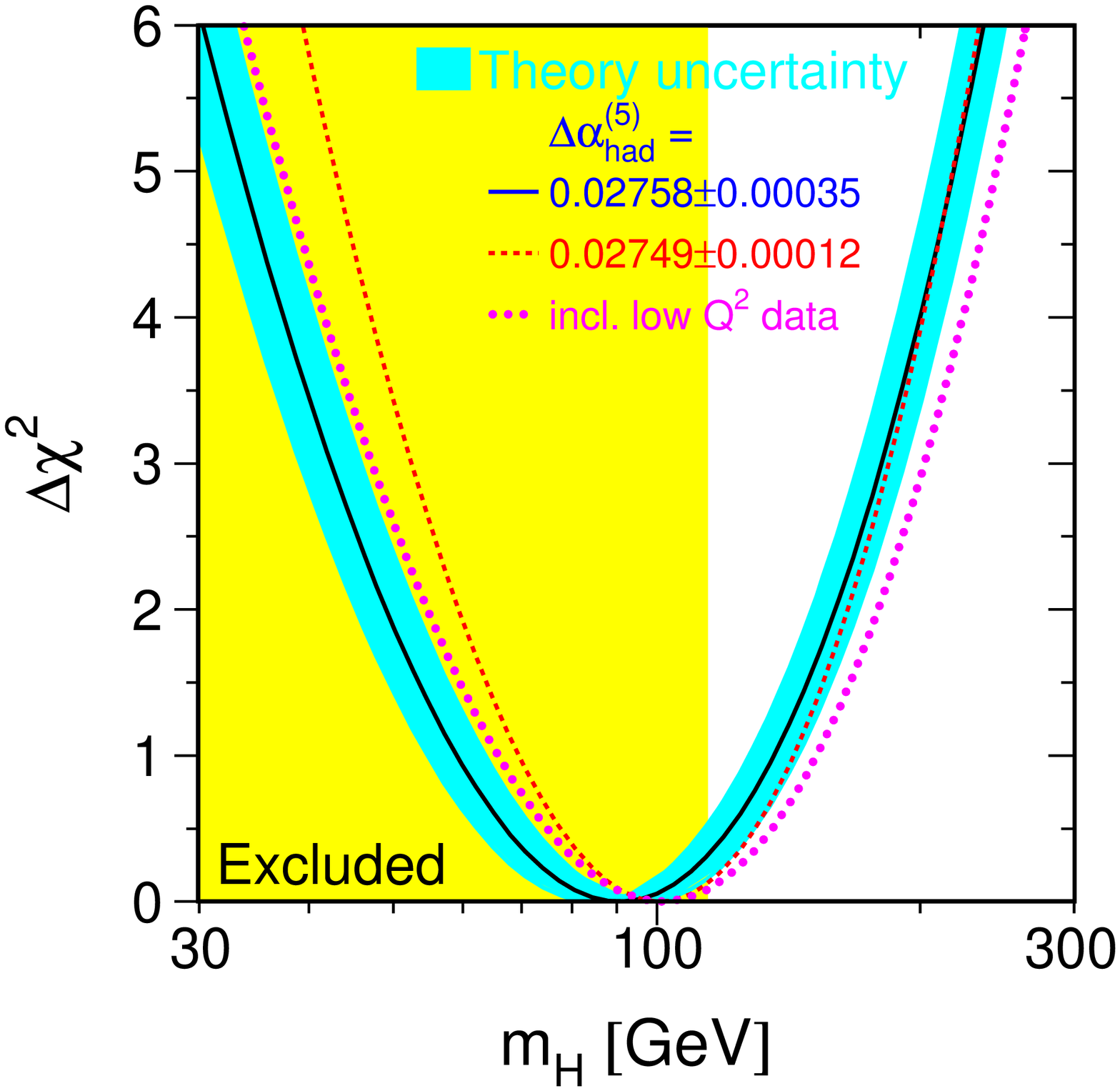}
\vskip -0.9cm
\caption{Left: Pulls of the measurements used in the global SM
  analysis. Right: $\Delta\chi^2$ curve as a function of $\MH$.  Also
  shown are the curves using a theory-driven evaluation of $\dalhad$,
  or including the low-$Q^2$ measurements.  }
\vskip -1cm
$ $
\label{fig:pulls-blue}
\end{center}
\end{figure}

The best constraint on $\MH$ is obtained by analysing all data.  This
global fit has a $\chi^2$ of 17.8 for 13 degrees of freedom,
corresponding to a probability of 16.6\%. The pulls of the 18
measurements entering the fit are shown in Figure~\ref{fig:pulls-blue}
(left).  The single largest contribution to the $\chi^2$ arises from
the $\Afbzb$ measurement discussed above, with a pull of 2.8.  The fit
yields $\MH = 91^{+45}_{-32}~\GeV$, which corresponds to a one-sided
95\% C.L. upper limit on $\MH$ of $186~\GeV$ including the theory
uncertainty as shown in Figure~\ref{fig:pulls-blue} (left).  The
fitted $\MH$ is strongly correlated with the fitted hadronic vacuum
polarisation (correlation of $-0.51$) and the fitted top-quark mass
($+0.52$).  The strong correlation with $\MT$ implies a shift of 20\%
in $\MH$ if the measured $\MT$ changes by $3~\GeV$ (about one standard
deviation).  Thus a precise measurement of $\MT$ is very important.
Also shown are the $\chi^2$ curves obtained with the more precise but
theory-driven evaluation of $\dalhad$~\cite{YNDURAIN}, yielding a
correlation of only $-0.2$ with $\MH$, or including the results
obtained in low-$Q^2$ interactions: atomic parity
violation~\cite{APV-Caesium}, Moller scattering~\cite{E-158}, and
NuTeV's measurement of deep-inelastic lepton-nucleon
scattering~\cite{NuTeV}; with the two former measurements in agreement
with the expectations but the latter differing by 3 standard
deviations.  Both analyses yield nearly the same upper limits on
$\MH$.

\clearpage

\clearpage

The theoretical uncertainty on the SM calculations of the observables
is visualised as the thickness of the blue band. It is dominated by
the theoretical uncertainty in the calculation of the effective
electroweak mixing angle, where a completed two-loop calculation is
needed.  The shaded part in Figure~\ref{fig:pulls-blue} (left) shows
the $\MH$ range up to $114.4~\GeV$ excluded by the direct search for
the Higgs boson at 95\% confidence level. Even though the minimum of
the $\chi^2$ curve lies in the excluded region, the uncertainties on
the Higgs mass value are such as that the results are well compatible.

\section{Conclusions}

\vskip -2mm

During the last 15 years many experiments have performed a wealth of
measurements with unprecedented precision in high-energy particle
physics. These measurements test all aspects of the SM of particle
physics, and many of them show large sensitivity to electroweak
radiative corrections at loop level.  Most measurements agree well
with the expectations as calculated within the framework of the SM,
successfully testing the SM at Born and at loop level. There are two
``3 standard deviations effects'', namely the spread in the various
determinations of the effective electroweak mixing angle, within the
SM analysis apparently disfavouring the measurement of $\Afbzb$, and
NuTeV's result, most pronounced when interpreted in terms of the
on-shell electroweak mixing angle.  For the future, precise
theoretical calculations including theoretical uncertainties are
needed, in particular a completed two-loop calculation for the
effective electroweak mixing angle.  Experimentally, the next few
years will bring improvements in the measurements of W and top masses,
and the long-awaited discovery of the Higgs boson.
\vfill
{\bf Acknowledgements}

It is a pleasure to thank my colleagues of the Tevatron and LEP
electroweak working groups, members of the E-158, NuTeV, SLD, ALEPH,
DELPHI, L3, OPAL, CDF and {D\O} experiments, as well as T.~Riemann and
G.~Weiglein for valuable discussions.


\begin{thebibliography}{99}
\bibitem{LEPEWWG}       LEP-EWWG, {\tt http://www.cern.ch/LEPEWWG}.
\bibitem{Z-POLE}        ALEPH, DELPHI, L3, OPAL, SLD, and the LEP-EWWG, 
                        hep-ex/0509008.
\bibitem{Mtop-EPS}      Koji Sato, these proceedings.
\bibitem{TEVEWWG05-EPS} CDF, D\O, and the TEV-EWWG, hep-ex/0507091. 
\bibitem{TEV-MW-GW}     CDF, D\O, and the TEV-EWWG, PRD 70 (2004) 092008.
\bibitem{MW-EPS}        Ambreesh Gupta, these proceedings;
                        Raimund Str\"ohmer, these proceedings.
\bibitem{BP05}          H.~Burkhardt, B. Pietrzyk, PRD 72 (2005) 057501.
\bibitem{TZ}            G.~Passarino \etal, CPC 117 (1999) 278;
                        D.~Bardin    \etal, CPC 133 (2001) 229.
\bibitem{YNDURAIN}      J.F.~de Troconiz, F.J.~Yndurain, 
                                          PRD 71 (2005) 073008
\bibitem{APV-Caesium}   J.~Ginges, V.~Flambaum, Phys. Rept. 397 (2004) 63.
\bibitem{E-158}         P.~Anthony \etal, E158 collaboration, 
                        PRL 95 (2005) 081601.
\bibitem{NuTeV}         G.P.~Zeller \etal, NuTeV collaboration, 
                        PRL 88 (2002) 091802, erratum 90 (2003) 239902.
\end{thebibliography}
\end{document}